\input harvmac
\input psfig
\newcount\figno
\figno=0
\def\fig#1#2#3{
\par\begingroup\parindent=0pt\leftskip=1cm\rightskip=1cm\parindent=0pt
\global\advance\figno by 1
\midinsert
\epsfxsize=#3
\centerline{\epsfbox{#2}}
\vskip 12pt
{\bf Fig. \the\figno:} #1\par
\endinsert\endgroup\par
}
\def\figlabel#1{\xdef#1{\the\figno}}
\def\encadremath#1{\vbox{\hrule\hbox{\vrule\kern8pt\vbox{\kern8pt
\hbox{$\displaystyle #1$}\kern8pt}
\kern8pt\vrule}\hrule}}
\def\underarrow#1{\vbox{\ialign{##\crcr$\hfil\displaystyle
 {#1}\hfil$\crcr\noalign{\kern1pt\nointerlineskip}$\longrightarrow$\crcr}}}
%
\overfullrule=0pt

%
\def\M{{\cal M}}
\def\tilde{\widetilde}
\def\bar{\overline}
\def\Z{{\bf Z}}
\def\T{{\bf T}}
\def\S{{\bf S}}
\def\R{{\bf R}}

\font\zfont = cmss10 

\def\bigone{\hbox{1\kern -.23em {\rm l}}}
\def\ZZ{\hbox{\zfont Z\kern-.4emZ}}

\Title{hep-th/9909229}
{\vbox{\centerline{Heterotic String Conformal Field Theory}
\bigskip
\centerline{And $A$-$D$-$E$ Singularities}}}
\smallskip
\centerline{Edward Witten}
\smallskip
\centerline{\it California Institute of Technology, Pasadena CA 91125 USA$^*$}
\smallskip
\centerline{\it and}
\smallskip
\centerline{\it CIT-USC Center For Theoretical Physics}
\bigskip

\medskip

\noindent

We analyze the behavior of the heterotic string near an $A$-$D$-$E$
singularity without small instantons.  This problem is governed
by a strongly coupled worldsheet conformal field theory, which, by
a combination of ${\cal O}(\alpha')$ corrections and worldsheet
instantons, smooths out the singularities present in the classical
geometry.

\vskip 6cm
\noindent
$^*$ On leave of absence from Institute for Advanced Study, Olden
Lane, Princeton NJ 08540.
\smallskip
\Date{October, 1999}

\def\K3{{\rm K3}}
\def\T{{\bf T}}
\newsec{Introduction}

String compactifications with  ${\cal N}=2$ supersymmetry in four
dimensions -- and more generally, with eight unbroken supercharges
in various dimensions -- have been much studied.  They are extremely rich 
in their behavior, yet sufficiently constrained
to be analyzed in detail.  

The low energy supergravity obtained in ${\cal N}=2$
compactification to four dimensions has
scalar fields in both vector multiplets and hypermultiplets.
The moduli space of vacua (endowed with the metric that appears in the low
energy effective action) is \ref\sugra{B. de Wit, P. Lauwers, and
A. Van Proeyen, ``Lagrangians Of $N=2$ Supergravity-Matter Systems,''
Nucl. Phys. {\bf B255} (1985) 269.} locally a product
of a vector multiplet moduli space and a hypermultiplet moduli space.

A prototype of such a compactification
 is the  heterotic string on $\K3\times \T^2$, which
is believed \nref\kachvafa{S. Kachru and C. Vafa, ``Exact Results For
$N=2$ Compactifications Of Heterotic Strings,'' Nucl. Phys. {\bf N450} (1995) 69, hep-th/9505105.}
\nref\ferr{S. Ferrara, J. A. Harvey, A. Strominger, and C. Vafa,
``Second-Quantized Mirror Symmetry,'' hep-th/9505162.}%
\refs{\kachvafa,\ferr}
to be dual to the Type IIA string on a Calabi-Yau
threefold.   (Which threefold arises here
depends on the heterotic string gauge bundle.)
In  compactification on $\K3\times \T^2$, 
the heterotic string dilaton is in
a vector multiplet, and the Type IIA dilaton is in a hypermultiplet.

As a result, the vector multiplet moduli space is 
independent of the Type
IIA string coupling, and the hypermultiplet moduli space is independent of
the heterotic string coupling.  Hence the vector multiplet moduli space
can  be determined, in principle, from Type IIA conformal field theory,
and likewise the hypermultiplet moduli space can be determined, in
principle, from heterotic string conformal field theory.  
\nref\more{S. Kachru, A. Klemm, W. Lerche, P. Mayr, and C. Vafa,
``Nonperturbative Results On The Point Particle Limit Of $N=2$
Heterotic String Compactifications,'' Nucl. Phys. {\bf B459} (1996) 537, hep-th/9508155.}%
This viewpoint has been much exploited for understanding
the vector multiplet moduli space  \refs{\kachvafa,\more}.
The hypermultiplet moduli space has also been much studied,
\nref\ostrom{A. Strominger, ``Loop Corrections To The Universal Hypermultiplet,'' Phys. Lett. {\bf B421} (1998) 139, hep-th/9706195.}
\nref\oasp{P. Aspinwall, ``Aspects Of The Hypermultiplet Moduli 
Space In String Duality,'' hep-th/9802194.}
for example in \refs{\ostrom,\oasp}, but is rather less understood.
The present paper will be devoted to some issues on
the hypermultiplet side, from the standpoint of heterotic string conformal
field theory.  

Since conformal field theory on $\K3 \times \T^2$
is the product of conformal field theory on K3 with (free) conformal
field theory on $\T^2$, the essential issues will involve the K3 conformal
field theory.  Thus, our problem will be to study the hypermultiplet
moduli space in compactification of the heterotic string on $\K3$.
For understanding the hypermultiplets, it does not matter much if
one considers K3 compactification to six dimensions or ${\rm K3}\times
\T^2$ compactification to four dimensions.

The claim that the moduli spaces can be computed from conformal
field theory is subject to an important caveat: the moduli spaces have
singularities, which sometimes reflect nonperturbative
physics -- like the massless hypermultiplet near a Type II conifold 
\ref\strom{A. Strominger, ``Massless Black Holes And Conifolds In
String Theory,'' Nucl. Phys. {\bf B451} (1995) 96, hep-th/9504090.}.
Thus, one can in principle compute the vector multiplet and
hypermultiplet moduli spaces using the appropriate conformal field
theories, but one may not be able to understand them.

An important example is the small instanton singularity of the heterotic
string.  The classical supergravity solution for a small instanton
\ref\chs{C. G. Callan, Jr., J. A. Harvey, and A. Strominger, ``World
Sheet Approach To Heterotic Instantons And Solitons,'' Nucl. Phys.
{\bf B359} (1991) 611.}
shows a blowup of the dilaton near the core of a small instanton,
so one must expect a nonperturbative phenomenon to occur as an instanton
shrinks to zero size.
The phenomenon in question is the appearance of a nonperturbative
gauge symmetry \ref\smallinst{E. Witten, ``Small Instantons In String
Theory,'' Nucl. Phys. {\bf B460} (1996) 541, hep-th/9511030.}.   Despite its importance in heterotic string dynamics,
the small instanton singularity is in the following sense not
 a good illustration
of the role of heterotic string conformal field theory.  The instanton
moduli space, as computed in classical field theory, has a small instanton
singularity, which is uncorrected in going to conformal field theory,
\foot{A proof that for $k$-instanton configurations on $\R^4$,
the classical instanton moduli space coincides with the Type I instanton
moduli space can be found in section 2.3 of \ref\noncomm{N. Seiberg and
E. Witten,  ``String Theory And Noncommutative Geometry,'' hep-th/9908142.}. 
Via heterotic - Type I duality, the same is therefore
also true for the heterotic string.} and is interpreted nonperturbatively
in terms of enhanced gauge symmetry.  Thus, heterotic string conformal
field theory (as opposed to supergravity) does not play an important role in
generating the singularity.  The interpretation of the singularity
is also out of reach of conformal field theory.
In the present paper, we will analyze an example in which heterotic string
conformal field theory does play a central role in controlling
the behavior near a classical singularity.

A hint about where to look
 comes from the classical equation of motion for the dilaton $\phi$, 
which reads schematically
\eqn\huggo{\bigtriangleup^2\phi =\tr\, F_{ij}F^{ij} -\tr\,R_{ij}R^{ij}.}
Here $\bigtriangleup^2$ is the Laplacian,
$F_{ij}$ is the Yang-Mills curvature of the gauge bundle, and $R_{ij}$
is the Riemann tensor (regarded as a two-form valued in the Lie algebra
 of the orthogonal group).   The crucial point is
the relative minus sign between the
two terms on the right hand side of this equation.  
If $F$ is large with $R$ zero, 
one is driven to strong coupling (as is familiar
from the small instanton solution \chs), while
if $R$ is large with $F$ zero, one is driven to weak coupling.  
So a singularity with large $R$ but zero $F$ should not lead to
nonperturbative physics, and should be understandable in the framework
of conformal field theory.

\def\R{{\bf R}}
We want a singularity that is at finite distance on the moduli space,
so we will concentrate on the $A$-$D$-$E$ singularities.  
$A$-$D$-$E$ singularities with small instantons have been much studied
and give interesting nonperturbative behavior 
\nref\aspinwall{P. Aspinwall and D. Morrison, ``Point-like Instantons
on K3 Orbifolds,'' hep-th/9705104.}%
\nref\intril{K. Intriligator, ``RG Fixed Points In Six Dimensions Via
Branes At Orbifold Singularities,'' hep-th/9702038.}%
\nref\bintril{J. Blum and K. Intriligator, ``Consistency Conditions For
Branes At Orbifold Singularities,'' hep-th/9705030, ``New Phases Of
String Theory And 6d RG Fixed Points Via Branes At Orbifold Singularities,''
hep-th/9705044.}  
\nref\donagi{P. Aspinwall and R. Donagi, ``The Heterotic String,
The Tangent Bundle, and Derived Categories,'' Adv. Theor. Math.
Phys. {\bf 2} (1998) 1041.}
\nref\mayr{P. Berglund and P. Mayr, ``Heterotic String/$F$-Theory Duality From Mirror Symmetry,'' hep-th/9811217.}
\refs{\aspinwall - \mayr}.  We will omit the small instantons
so as to get an example governed by conformal field theory.
Near the singularity, one can
replace the ambient K3 manifold by an ALE space  (which
is asymptotic to $\R^4/\Gamma$
for some finite group $\Gamma$; $\Gamma$ depends on the choice of
an $A$-$D$-$E$ singularity).  Since there are no small instantons,
we can set $F=0$. 
So we will study the heterotic string
on an ALE space with $F=0$, in the conformal field theory limit.
\foot{Because we work at string tree level, we will not see the
fluctuations around $F=0$, which would (in string loops) distinguish
the $E_8\times E_8$ and ${\rm Spin}(32)/\Z_2$ heterotic strings.
Thus, the two heterotic string theories, compactified on an ALE space without
small instantons, have the same moduli space. Note that in the case
of ${\rm Spin}(32)/\Z_2$, since we have taken the gauge field
to be trivial, we are considering the case of  a gauge bundle with
vector structure.}

The conformal field theory moduli space ${\cal M}$
for the heterotic string on
such a manifold is a hyper-Kahler manifold of dimension
$4r$, where $r$ is the rank of the $A$-$D$-$E$ group in question.
That ${\cal M}$ is hyper-Kahler requires some explanation.  In general,
the hypermultiplet moduli space in a globally supersymmetric theory
with eight unbroken supercharges is a hyper-Kahler manifold,
but in the presence of gravity it is instead a quaternionic
manifold.  Focussing on the behavior near
a singularity has the effect of decoupling gravity, and that is
why ${\cal M}$ is hyper-Kahler.

We will focus on the simplest case of $A_1=SU(2)$, so that $r=1$ and
${\cal M}$ has real dimension four.  In this case, we will in section
2 analyze
the structure of ${\cal M}$ in the following three steps:

\def\Z{{\bf Z}}
\def\R{{\bf R}}
\def\S{{\bf S}}
(1) In supergravity, one can compute directly that ${\cal M}=(\R^3\times
\S^1)/\Z_2$, where the $\Z_2$ acts by multiplication by $-1$ on
both factors.  

(2) Going to conformal field theory, there is an ${\cal O}(\alpha')$
correction with the following structure.  The correction is singular
at the origin (the $\Z_2$ fixed point) in $\R^3$.  Let $\tilde\R^3$
be $\R^3$ with the origin deleted.  Then the ${\cal O}(\alpha')$ 
correction has the effect of replacing $\tilde \R^3\times \S^1$
by a twisted $\S^1$ bundle over $\tilde\R^3$ (which must then be
divided by $\Z_2$ to get the moduli space).  Since $\tilde \R^3$
is homotopic to $\S^2$, such bundles are classified by an integer-valued
first Chern class.  In this case, the integer is equal to $-4$.

(3) There are no further worldsheet perturbative corrections to the metric
of ${\cal M}$; that is, there are no corrections of order $(\alpha')^s$
with $s>1$.  However \ref\witteno{E. Witten, ``World-Sheet Corrections
Via $D$-Instantons,''World-Sheet Corrections Via $D$-Instantons,''
hep-th/9907041.}, there are worldsheet instanton corrections
to ${\cal M}$.  These corrections will preserve the hyper-Kahler structure
of ${\cal M}$ as well as an $SO(3)$ action on ${\cal M}$ that rotates
the complex structures and whose generic orbits are three-dimensional.  Moreover, the worldsheet instanton corrections vanish at infinity (on $\R^3$),
and do not modify
the asymptotic behavior of ${\cal M}$ as found in step (2) above.
Four-dimensional hyper-Kahler manifolds with these properties have
been analyzed 
\nref\gibbons{G. Gibbons and C. Pope, ``The Positive Action Conjecture
And Asmptotically Euclidean Metrics In Quantum Gravity,'' Commun. Math.
Phys. {\bf 66} (1979) 267.}%
\nref\ah{M. F. Atiyah and N. Hitchin,
{\it The Geometry And Dynamics Of Magnetic Monopoles} (Princeton University
Press, 1988).}%
\refs{\gibbons,\ah}.
Such an ${\cal M}$ either has a singularity that would be difficult to
interpret in conformal field theory, or is a unique, smooth,
complete hyper-Kahler four-manifold ${\cal M}_{AH}$ described in \ah.  
We thus argue that ${\cal M}={\cal M}_{AH}$.  Here ${\cal M}_{AH}$ is
the space that was identified by Atiyah and Hitchin as the moduli space of
BPS dimonopoles in three dimensions.

In sections 2.3 and 2.4, we give the best arguments we can for why
the moduli space should be nonsingular.  Among other things, we argue
by a simple linear sigma model construction that the $(0,2)$ conformal
field theory describing the heterotic string on a Calabi-Yau manifold $Y$
of any complex dimension $n$ never develops a singularity when $Y$ develops
an isolated hypersurface singularity near which the gauge fields
are trivial.  


Steps (1) to (3)
above are in precise parallel with similar steps that were used
in the determination of the moduli space of vacua for minimal $SU(2)$ 
supersymmetric gauge theory in three dimensions with eight supercharges
\ref\seiwit{N. Seiberg and E. Witten, ``Gauge Dynamics And Compactification
To Three Dimensions,'' in {\it The Mathematical Beauty of Physics},
hep-th/9607163.}.  (Minimal means that we consider
the theory of the $SU(2)$ vector multiplet only, without additional
charged fields.)
In that case (after dualizing
the photon to convert the vector multiplet that parametrizes the Coulomb
branch of the theory to a hypermultiplet), we have the following
statements, in close parallel to the above: (1)  
the classical moduli space of vacua is $(\R^3\times \S^1)/\Z_2$;
(2) there is a one-loop correction that replaces the product $\R^3\times
\S^1$ by
a twisted fiber bundle;  (3) there are instanton corrections that
turn the moduli space into ${\cal M}_{AH}$.

This analogy
suggests a generalization of our result to other $A$-$D$-$E$
singularities.  Consider the heterotic string at a singularity of
type $G$ without small instantons, where $G$ is a group of $A$, $D$, or
$E$ type.  The conjecture is that the hypermultiplet moduli
space for the heterotic string near such a singularity is
the moduli space of vacua of a minimal supersymmetric gauge theory
in three dimensions with eight supercharges and gauge group $G$.

After submitting to hep-th the original version of this paper,
I became aware that Sen has treated the $A_n$ case of this problem, by considering the heterotic string on
a multi-Taub-NUT spacetime \ref\sen{A. Sen, ``Dynamics Of Multiple
Kaluza-Klein Monopoles In $M$ And String Theory,'' Adv. Theor. Math. Phys.
{\bf 1} (1998) 115.}.  (The multi-Taub-NUT example is relevant because
it can develop an $A_n$ singularity.)
In this approach, the connection to BPS multimonopoles of $SU(2)$ gauge
theory is made by going very close to the self-dual radius of the heterotic
string on a circle, where an enhanced $SU(2)$ gauge symmetry appears.
See also \ref\blum{J. Blum, ``$H$ Dyons And $S$ Duality,'' Nucl. Phys. {\bf B507} (1997) 245, hep-th/9702084.} for a prior discussion of the relation of
$H$-monopoles to BPS monopoles.

\newsec{Analysis Of The Moduli Space}

As was explained in the introduction, we will here
analyze the behavior of the heterotic string at an $A_1$ singularity
without small instantons.  An $A_1$ singularity is simply a quotient
singularity of the form $\R^4/\Z_2$, where the generator of $\Z_2$
acts on $\R^4$ by multiplication by $-1$.
The analysis will come in the three stages described in the introduction:
(1) supergravity; (2) incorporation of an ${\cal O}(\alpha')$ correction;
(3) exact description using worldsheet instantons.

\subsec{Supergravity Analysis}

In string theory, 
when one divides $\R^4$ by $\Z_2$ to form the orbifold $\R^4/\Z_2$,
one must pick the sign of the action of $\Z_2$ on fermions.
Either choice leaves half of the supersymmetry unbroken and determines
a distinguished orientation on $\R^4/\Z_2$.  With this distinguished
orientation, $\R^4/\Z_2$ is a flat hyper-Kahler manifold with an isolated
orbifold singularity.  As a hyper-Kahler manifold, $X$ has a two-sphere
of complex structures.  If $I$, $J$, and $K$ are the quaternion generators
on $\R^4$, then the general complex structure is $w_1I+w_2J+w_3K$,
where $w_1^2+w_2^2+w_3^2=1$.  The symmetry group of $X=\R^4/\Z_2$ is
$SO(4)=SU(2)_L\times SU(2)_R$ where (with a suitable choice of orientation)
$I,J,$ and $K$ are invariant under $SU(2)_L$ and transform with spin one
under $SU(2)_R$.  Hence a choice of $\vec w$ breaks $SU(2)_L\times SU(2)_R$
to $SU(2)_L\times U(1)_R$.

If one picks a particular $\vec w$, that is a particular
complex structure on $X$, 
then one can ``blow up'' the orbifold singularity of $X$ 
in that complex structure to make a smooth ALE
hyper-Kahler manifold, the Eguchi-Hansen space $X'$
\ref\eguchi{T. Eguchi and A. J. Hanson, ``Asymptotically Flat
Self-Dual Solutions To Euclidean Gravity,'' Phys. Lett. {\bf 74B}
(1978) 249.}.  In such a blowup, there is a projection $X'\to X$
which is generically one-to-one and is holomorphic in the complex structure
specified by $\vec w$.   
The blowup of $X$ to make $X'$ is completely determined by the
choice of  $\vec w$ and the area $\alpha$ of the
exceptional divisor produced in the blowup.  We can combine $\vec w$ and
$\alpha$ to a three-vector
$\vec m$ whose direction is the unit vector $\vec w$ and
whose magnitude determines $\alpha$. (A convenient way to do this
is described below.)  
$\vec m$ is a completely arbitrary element of $\R^3$, so at first sight
it seems that the moduli space of hyper-Kahler
 blowups of $X$ (modulo diffeomorphisms
that are trivial at infinity) is a copy of $\R^3$.

However, the complex structure $\vec w$ with respect to which the blowup
is made is not quite uniquely determined.
The projection $X'\to X$ is holomorphic
with respect to both the complex structure determined by $\vec w$, and
the ``opposite'' (or complex conjugate) complex structure determined by
$-\vec w$.  Hence, we should identify $\vec m$ with $-\vec m$, and
the moduli space of blowups is actually  $\R^3/\Z_2$.

We can verify this in the following direct way.  As originally
presented in \eguchi, the metric of $X'$ reads
\eqn\dolgo{ds^2=f(r)^2dr^2+r^2(\sigma_1^2+\sigma_2^2)+r^2g^2\sigma_3^2,}
with
\eqn\olgo{g=f^{-1}=\sqrt{1-(a/r)^4};}
here 
$\sigma_1,\sigma_2$, and $\sigma_3$ are the left-invariant one-forms
on $\S^3\cong SU(2)$, and $a$ is a constant that can be identfied
with $|\vec m|$.   Note that the distinguished role of $\sigma_3$
in the formula breaks the $SU(2)_L\times SU(2)_R$ symmetry of
the $SU(2)$ manifold to $SU(2)_L\times U(1)_R$, as expected.
We can readily generalize \dolgo\ to a more general blowup in which
$\sigma_3$ is replaced by a more general linear combination of the 
$\sigma$'s.  We introduce an arbitrary unit vector
$\vec w$ and generalize \dolgo\ to
\eqn\ufolgo{ds^2=f(r)^2dr^2+r^2\left(\sum_{i=1}^3\sigma_i^2
-(\vec w\cdot \vec\sigma)^2\right) +r^2g(r)^2(\vec w\cdot 
\vec\sigma)^2.}
This is invariant under $\vec w\to -\vec w$, so we should consider
$\vec w$, and hence also $\vec m=a\vec w$, to be defined only
up to sign.  

The exceptional divisor $S$ produced in the blowup is the two-sphere
at $r=a$ (where the coefficient of $\sigma_3$ in \dolgo\ vanishes
and the $\S^3$ collapses to an $\S^2$).  Its area is $\alpha=4\pi a^2$.

Now, let us include the $B$-field.  This introduces one more real modulus,
which is essentially
the period of the $B$-field integrated over $S$:
\eqn\ikko{\theta=\int_SB.}
Modulo global gauge transformations of the $B$ field, $\theta$ is
an angular variable, of period $2\pi$.  At first sight, then, it seems
that the supergravity moduli space of $X'$ is a product
$\R^3/\Z_2\times \S^1$, where the first factor allows for the blowup
and the second for the theta angle.   But $\R^3/\Z_2\times\S^1$ is not
hyper-Kahler, so inevitably there is a subtlety here.  To make
sense of $\theta$ as a number, we need an orientation of $S$.  While
$X'$ has a natural orientation, $S$ does not.  A choice of complex structure
on $X'$ determines a complex structure and hence an orientation of $S$,
but if we reverse the complex structure of $X'$, the complex structure
and orientation  of $S$ will be reversed.

So under $\vec w\to -\vec w$, the orientation of $S$ is reversed and
$\theta$ is mapped to $-\theta$.  The supergravity moduli space of
$X'$ is thus
\eqn\polco{{\cal M}_{SUGRA}=\left(\R^3\times \S^1\right)/\Z_2,}
with $\Z_2$ acting as $-1$ on both $\R^3$ and $\S^1$.  This carries
a natural flat hyper-Kahler metric, with two isolated orbifold singularities.

What about the symmetries of $\M_{SUGRA}$?  $SU(2)_L$ acts
trivially on $\M_{SUGRA}$, since it is left unbroken by the blowup,
regardless of the choice of $\vec w$, and hence acts trivially on $\vec w$.
But $SU(2)_R$ acts on $\M_{SUGRA}=(\R^3\times \S^1)/\Z_2$ as
 the group of rotations of $\R^3$.  It rotates the complex structures
of $\M_{SUGRA}$ just as it did to the original orbifold $X$.
The double cover $\R^3\times \S^1$ also has a $U(1)$ symmetry,
which we will call $U(1)_A$, that rotates the $\S^1$ factor,
adding a constant to the period of the $B$-field.    $SU(2)_R$, because
it originates in the symmetries of the original orbifold $X$, is an exact
symmetry of the conformal field theory moduli space, but $U(1)_A$ is
broken by worldsheet instantons.

\bigskip\noindent{\it Topology}

\def\O{{\cal O}}
To compute the ${\cal O}(\alpha')$ correction to the moduli
space in section 2.2,
we will need some information about the topology of $X'$, and in more
detail about the behavior of $X'$ as $\vec w$ varies.

For fixed and nonzero $\vec w$, $X'$ is the cotangent bundle
$T^*S$ of the exceptional divisor $S$.  This is a standard fact, which
we review below.  For the present discussion,
we will mainly limit ourselves to the case of a unit blowup,
$|\vec w|=1$; the reason for this is that as long as the area of the
exceptional divisor is nonzero, it can be scaled out and does not affect
the topology.  We will also mostly ignore the discrete identification
$\vec w \to -\vec w$, and consider $\vec w$ to parametrize a two-sphere
$W$.  At the end of the discussion, one can divide by $\Z_2$.

As $\vec w$ varies in $W$, $X'$ varies as the fibers of a six-manifold
$Y$ that is fibered over $W$.  $Y$ is not a simple product $X'\times W$.
However, if we restrict ourselves to the exceptional divisor,
we do get a simple product $S\times W$.  This is ensured by the
$SU(2)_L\times SU(2)_R$ group action, with $SU(2)_L$ and $SU(2)_R$
acting, respectively, by rotations of $S$ and $W$.  Replacing $S\times W$
by a nontrivial fibration would spoil the $SU(2)_L\times SU(2)_R$
symmetry.  Thus $Y$ contains an embedded copy $S\times W$ of
$\S^2\times \S^2$.  The normal bundle $N$ to $S\times W$ in $Y$ is
a real two-plane bundle, which we can alternatively regard
as a complex line bundle ${\cal R}$.  A complex line bundle is
labeled topologically by its first Chern class.
In the present case, the first Chern class of ${\cal R}$ takes values in 
$H^2(S\times W;\Z)=\Z\times \Z$, and is determined by a pair
of integers which are the components of the first Chern class along
$S$ and $W$, respectively.\foot{To interpret these components
as integers requires orienting $S$ and $W$.  We will not be precise
about the orientations, so some of our statements only hold
up to sign.}  We will show that these integers
are $(2,2)$, a fact that will be used in section 2.2 to determine
the $\O(\alpha')$ correction.   

To show this, we will use an alternative description of $X$ and $X'$
as a hyper-Kahler quotients \nref\kron{P. Kronheimer,  ``The Construction
Of ALE Spaces As Hyper-Kahler Quotients,'' J. Diff. Geom. {\bf 28} (1989)
665.}%
\nref\dougmoore{M. Douglas and G. Moore,  ``$D$-Branes, Quivers, and ALE
Instantons,'' hep-th/9603167.}%
\refs{\kron,\dougmoore}.
  We let $a^{AA'}$, $A,A'=1,2$ be a complex
hypermultiplet; here $A$ and $A'$ transform as spin 1/2  of $SU(2)_L$
and $SU(2)_R$, respectively, and in addition $a^{AA'}$ has charge
1 with respect to a $U(1)$ gauge group that we will call $U(1)_G$.
The hyper-Kahler moment map condition (for a supersymmetric vacuum
after gauging of $U(1)_G$) IS 
\eqn\tolgo{\sum_A a^{AA'}\bar a_{AB'}=\vec w\cdot \vec \sigma^{A'}{}_{B'},}
with $\vec \sigma$ the Pauli $\sigma$-matrices.  After imposing this
condition and dividing by $U(1)_G$, one gets for $\vec w=0$ the
orbifold $X=\R^4/\Z_2$, and for $\vec w\not= 0$ the resolution $X'$.

Let explicitly $a^{AA'}=u^A\delta^{A'1}+v^A\delta^{A'2}$.
The hyper-Kahler moment map equation is in more detail
\eqn\ufolgo{\eqalign{\sum_A\left(|u^A|^2-|v^A|^2\right) = & w_3\cr
                     \sum_Au^A\bar v_A  =& w_1+iw_2.\cr}}
For example, suppose $w_1=w_2=0$, $w_3=1$.  If we set $v^A=0$
and divide by $U(1)_G$, we get a copy of ${\bf CP}^1$ which is the
exceptional divisor $S$.  Relaxing the condition $v^A=0$, the
equation $\sum_Au^A\bar v_A=0$ shows that $v$ is a cotangent vector
to $S={\bf CP}^1$, so that $X'$ is the cotangent bundle of $S$,
as mentioned earlier.
  
Now we want to describe the normal bundle $N$, or equivalently
the complex line bundle ${\cal R}$.  Since ${\cal R}$ is a homogeneous
($SU(2)_L\times SU(2)_R$-invariant) line bundle over the homogeneous
space $S\times W$, it can be uniquely determined by describing the
group action.  Picking a point 
$P$ (such as the point where $w_1, w_2$, and $u^2$ all vanish) in
$S\times W$, $SU(2)_L\times SU(2)_R$ is broken down to
$U(1)_L\times U(1)_R$.  $U(1)_L\times U(1)_R$ acts on the fiber
of ${\cal L}$ at $P$ with some charges, say $(n,m)$, and these
are the components of the first Chern class.   We will
now show that the charges are $(2,2)$.

In fact, $U(1)_L$ is the symmetry under which $u^1$ and $v^1$
have charge 1 and $u^2$ and $v^2$ have charge $-1$, while $U(1)_R$
is the symmetry that assigns charge 1 to $u^A$ and $-1$ to $v^A$.
The $U(1)_G$-invariant coordinate on the fiber of ${\cal L}$ over $P$
is $q=u^1\bar v_2$, which has charge 2 for both $U(1)_L$ and $U(1)_R$,
as promised.

To conclude, we will tie up a detail.  Using the description
in \tolgo, the identification $\vec w\leftrightarrow -\vec w$ is
not very evident.   This identification arises because the
transformation $\tau:a^{AA'}\leftrightarrow \epsilon^{AB}\epsilon^{A'B'}
\bar a_{BB'}$ maps $\vec w\to -\vec w$, so resolutions of the singularity
with equal and opposite $\vec w$ are equivalent by the action of $\tau$.
Moreover (after imposing \tolgo\ and dividing by $U(1)_G$),      $\tau$
acts trivially for $|a|\to\infty$.  The last condition is important,
because in describing the moduli of $X'$, we classify the resolutions
up to diffeomorphisms that are trivial near infinity; all of the
deformations with the same $|\vec w|$ 
are in fact equivalent by an $SU(2)_R$ rotation  which acts
nontrivially at infinity.

\subsec{The ${\cal O}(\alpha')$ Correction}

In this subsection, we carry out the second step in analyzing the
heterotic string moduli space on the ALE space without small 
instantons.  This is to analyze
the corrections to the moduli space coming from worldsheet perturbation
theory.  We will find that the perturbative correction to the metric
is completely determined by an ${\cal O}(\alpha')$ term that
can be described in terms of topology.

\def\TT{{\cal T}}
The supergravity moduli space $\M_{SUGRA}=
(\R^3\times \S^1)/\Z_2$ that we found
in \polco\ is an $\S^1$ bundle over
${\cal W}=\R^3/\Z_2$, where the fibration is described by forgetting
$\theta$.  This fibration is flat: it is trivial if lifted
to $\R^3$ (since the double cover of ${\cal M}_{SUGRA}$ is a product
$\R^3\times \S^1$).  For Type II superstrings, something like this is
 the complete answer to all orders in $\alpha'$:     
the $B$-field periods take values in a torus
bundle (or a circle bundle when there is only one period,
as in the case that we are studying) that is flat, with discrete monodromies
(which are associated with the mapping class group and singularities).
For the heterotic string, however, the $B$-field periods take values
in a circle or torus bundle $\TT$ that is not flat.  (In addition, as we will
see, the fibration structure breaks down near certain singularities.) 

Note that a circle bundle ${\cal V}$ is closely related to a complex
line bundle ${\cal V}'$; ${\cal V}$ is the bundle of unit vectors in 
${\cal V}'$.  We will write $c_1({\cal V})$ as an abbreviation for
$c_1({\cal V}')$.  A torus bundle ${\cal T}$ is similarly related
to a bundle whose fiber is ${\bf C}^r$, with $r$ the rank of the torus.

Here are two related approaches to analyzing the torus bundle
$\TT$:

(1) For Type II, the field strength of $B$ is $H=dB$, and the Bianchi
identity reads
\eqn\biid{dH=0.}
For the heterotic strings, there are additional Chern-Simons terms
in the definition of $H$, and the Bianchi identity reads
$dH=(\tr F\wedge R-\tr R\wedge R)/4\pi$, where $F$ and $R$ are the
Yang-Mills and Riemann curvature two-forms.  In the present paper, we set
$F=0$, so the Bianchi identity (after dividing by $2\pi$ for convenience
since $H/2\pi$
has integral periods) is
\eqn\niid{d\left({H\over 2\pi}\right)=-{1\over 8\pi^2}\tr\,R\wedge R.}
The right hand side comes from an $\O(\alpha')$ correction in the
worldsheet theory (though we have set $\alpha'=1/2\pi$ in writing
the formula).    
For a given target space $X'$ of the heterotic string, \niid\ has solutions
(or the model would be altogether inconsistent).  Now suppose
that $X'$ varies in its moduli space.  If there exists a smoothly
varying solution $H_0$ of \niid, then one can set $H=H_0+dB'$, where
$B'$ is an ``ordinary two-form gauge field'' (whose field strength,
for example, obeys conventional Dirac quantization); by taking the periods
of $B'$ as coordinates, this would trivialize $\TT$.
The obstruction to trivializing $\TT$ is thus the obstruction
to picking a smoothly varying $H_0$.  
This gives us a framework for describing $\TT$; if the $B$-field has
only one period $\int_SB$, so $\TT$ is a circle bundle, then
\eqn\uncup{\int_Wc_1({\cal T})=\int_{S\times W}\left(-{1\over 8\pi^2}
\tr\, R\wedge R\right)}
 for any two-cycle $W$ in the moduli space.
\foot{We are over-simplifying a bit.
In general, in the above formula, $S$ varies with $W$, and $S\times 
 W$ must be replaced by the total space of a fiber bundle,
with fiber $S$ and base   $W$.}
 This formula
completely determines $c_1({\cal T})$ modulo torsion. 
In the present case there
is only one relevant two-cycle and no possibility of torsion.
There is a natural parallel
transport of $H_0$ that comes by asking that its change (when the
metric of $X'$ and hence the right hand side of \niid\ varies) be
as small as possible; in going around a loop in moduli space, $H_0$ does
not come back to itself, which is why ${\cal T}$ is not flat.
Obviously, if ${\cal T}$ were flat, then the left hand side of \uncup\ would
vanish.

\def\Pf{{\rm Pf}}
\def\D{{\cal D}}

(2) A related and more precise approach (which, for example, could
be used in a more complicated situation to determine the torsion
in $c_1({\cal T})$) is as follows.  Suppose
that we want to study the period of $B$ integrated over a two-cycle $S$
in spacetime.  One factor in the worldsheet path integral is the 
coupling $\exp(i\int_SB)$ to the $B$-field; another factor is the
Pfaffian $\Pf(\D)$ of the worldsheet Dirac operator $\D$.  The product
\eqn\olop{\exp(i\int_SB)\,\cdot \Pf(\D)  }
must be well-defined.  The Pfaffian $\Pf(\D)$ takes values in
a ``Pfaffian line bundle'' ${\cal L}$.  Hence, 
$\exp(i\int_SB)$ must be a section of ${\cal L}^{-1}$.  So
the period $\int_SB$ of the $B$-field does not take values in 
$\R/2\pi \Z$ but in a circle bundle ${\cal T}$ which is the bundle
of unit vectors in ${\cal L}^{-1}$.  (This description of ${\cal T}$,
which is discussed in more detail in \witteno, can
be reduced to the previous one by using the Quillen formula for the
curvature of ${\rm Pf}(\D)$.)

In our problem of strings on the ALE space $X'$, the moduli space of
hyper-Kahler metrics is $\R^3/\Z_2$.  At the origin in $\R^3/\Z_2$,
$X'$ develops a singularity and the $\alpha'$ expansion breaks down, as do
the definition and interpretation of the $B$-field
period.  Hence, in analyzing the $\O(\alpha')$ correction, we will
work away from the origin in $\R^3/\Z_2$.  As in the discussion at the
end of section 2.1, this means for topological purposes that we can
replace $\R^3/\Z_2$ with the unit sphere $W$ defined by $|\vec w|=1$;
 $W$ should be
divided by $\Z_2$ at the end of the discussion.

In the particular case that we are looking at, the 
$B$-field has
only one period, namely $\theta=\int_SB$ with $S$ the exceptional divisor.
This period takes values in a circle bundle ${\cal T}$ over $W$; we wish
to compute the first Chern class of ${\cal T}$.  We will do this using
the approaches (1) and (2) above:

$(1)'$ The characteristic class $-\tr R\wedge R/8\pi^2$ that appears in
the Bianchi identity \niid\ is $-p_1/2$, where $p_1$ is the first
Pontryagin class.  As in section 2.1, let $Y$ be the six-manifold
fibered over $W$ with fiber $X'$ (the fiber over $w\in W$ being
$X'$ with moduli determined by $\vec w$).  \uncup\ amounts
to
\eqn\juncup{\int_Wc_1({\cal T})=-{1\over 2}\int_{S\times W}p_1(TX'),}
with $TX'$ the tangent bundle of $X'$. (It would not matter if we used
the tangent bundle of $Y$ instead.)
 In section 2.1, we showed
that $Y$ is fibered over $S\times W$
with normal bundle a two-plane bundle $N$ or equivalently a complex
line bundle ${\cal R}$.  Because $Y$ is contractible to $S\times W$, 
the class
$-p_1(TX')/2$ is a pullback from $S\times W$.  To evaluate it, we note
that $TX'$ 
restricted to $S\times W$
is $TS\oplus N$ (where $TS$ is the tangent
bundle of $S$).  As $TS$ is stably trivial, it
does not contribute to $p_1(TX')$, which hence receives a contribution
only from $N$.  In general, one has for any real vector bundle $Q$, $p_1(Q) =\sum_ix_i^2$, where the
$x_i$ are the roots of the Chern polynomial.
For $Q$ a two-plane bundle $N$ that is associated with a complex
line bundle ${\cal R}$,
  there is only one root, which is
$c_1({\cal R})$.  We computed this in section 2.1 to be $2[S]+2[W]$,
where the intersection numbers are $[S]^2=[W]^2=0$, $[S]\cdot [W]=1$.
So $-\int_{S\times W}p_1/2=-(1/2)\int_{S\times W}c_1({\cal R})^2=-(1/2)(2[S]+2[W])^2=-4$.
The first Chern class of the line bundle $\TT$ over $W$ is thus $-4$.

\def\L{{\cal L}}
$(2)'$ For the second approach, we take $S$ in \olop\ to be the exceptional
divisor, and we must identify the Pfaffian line bundle $\L$ as a line
bundle over $W$.  $W$ is a two-sphere that is a homogeneous space for
$SU(2)_R$, and a given point $w\in W$ is invariant under a subgroup
$U(1)_R$ of $SU(2)_R$.  The first Chern class of $\L$ is simply
the ``charge'' (or the eigenvalue of the generator) with which $U(1)_R$
acts on the fiber of $\L$ over $w$.  But this fiber is simply the top
exterior power of the
space of zero modes of the worldsheet fermions of the heterotic  string,
with worldsheet $S$.  So the charge of the fiber is the sum of the
charges of the zero modes.
The heterotic string worldsheet fermions are
left-moving gauge fermions $\lambda$, which in our problem have no zero modes
since we have taken the gauge fields to vanish, and right-moving spacetime
fermions $\psi$, which are spinors on $S$ with values in the tangent
bundle to $X'$.  The only modes of $\psi$ that matter for computing
the $U(1)_R$ action on the fiber of ${\cal L}$ are the modes that
transform nontrivially under $U(1)_R$; these are the modes that are sections
of the normal bundle $N$ (to $S$ in $X'$).  As we explained in
section 2.1, $N$ is the cotangent bundle to $S$, rotated with charge 2
by $U(1)_R$.  Holomorphically, the spin bundle of $S$ is the holomorphic
bundle $\O(-1)$.  $N$ is the real cotangent bundle of 
$S$; its complexification
splits holomorphically
as $\O(-2)^{-2}\oplus \O(2)^{2}$, where the exponent is the $U(1)_R$
charge.  Tensoring this with the spin bundle ${\cal O}(-1)$, it follows
that spinors on $S$ with values in $N$ are  the sum of $\O(-3)^{-2}$,
with no zero modes, and $\O(1)^{2}$, with two zero modes.  Since
these two zero modes each have charge $2$, the total $U(1)$ charge of the
zero modes is $2+2=4$.  The first Chern class
of $\L$ is hence $4$, and so the inverse bundle $\TT$, of which the
$B$-field period is a section, has first Chern class $-4$.

\bigskip\noindent{\it The Metric}

The reader may be perplexed: our goal was to compute the string
perturbation theory
corrections to the metric on the moduli space ${\cal M}$, 
and instead we have computed the first Chern class of a line bundle.

There is, however, a simple relation between the two questions.
As noted in section 2.1,  the supergravity moduli space, before
dividing by the discrete symmetry $\tau$, 
has a $U(1)_A$ symmetry that rotates the period of $B$ by
$\theta\to \theta+c$ for any angle $c$.  This is also a symmetry
of the $\alpha'$ expansion, since the zero mode of $B$ decouples
in sigma model perturbation theory.
The metric on ${\cal M}$ computed in sigma model perturbation
theory therefore has this $U(1)$ symmetry.
It also, of course, has $SU(2)_R$ symmetry, induced from the geometric
symmetries of $X$ that are broken by the blowup.  It acts with generic
orbits three-dimensional, since the first Chern class at infinity
is nonzero.  (If indeed $SU(2)_R$ acted only on the base and not the
fiber of the fibration at infinity, then the $SU(2)_R$ orbits would
give a trivialization of that fibration.) 

Hyper-Kahler metrics in four dimensions with this kind of $SU(2)\times U(1)$
symmetry have been classified \ah\ and are
completely determined by the topology at infinity.  Such metrics
are constructed from the Euclidean
Taub-NUT space.  This space can be obtained by a hyper-Kahler
quotient \ref\ggibbons{G. Gibbons and P. Rychenkova, ``Hyper-Kahler Quotient
Construction Of BPS Monopole Moduli Spaces,'' hep-th/9608085.} 
and can be explicitly described by the metric
\eqn\zzzz{ds^2={1\over 4}\left({1\over |\vec x|}+{1\over \lambda^2}\right)
d\vec x^2+{1\over 4}\left({1\over |\vec x|}+{1\over \lambda^2}\right)^{-1}
\left(d\theta+\vec\omega\cdot d\vec x\right)^2,}
with $\lambda$ a constant and $\vec \omega$ the Dirac monopole
potential on ${\bf R}^3$.  
This manifold is smooth and has the topology of $\R^4$; 
the group $U(1)_A$ of shifts
of $\theta$ has a fixed point at the origin.  The fixed point means
that at the origin, there is no such thing as ``the period of the $B$-field.'' 
At infinity, the Taub-NUT space
looks like an $\S^1$ bundle over $\R^3$  of first Chern class $-1$.
To get first Chern
class $-n$ at infinity, with $n>0$, still over $\R^3$, one
    divides by $\theta\to \theta+2\pi/n$,
producing a $\Z_n$ orbifold singularity at the origin.  (First Chern
class $+n$ at infinity with $n>0$ is obtained from the same metric
with opposite orientation of the fiber.)  In our case, $n=4$, and
we want the structure at infinity to be that of an $\S^1$ bundle
over $\R^3/\Z_2$, not $\R^3$, so we must divide by an additional $\Z_2$.
The generator of this $\Z_2$ acts on $\theta$ by $\theta\to -\theta$;
this transformation
together with the $\Z_4$ symmetry $\theta\to \theta+\pi/2$
generates   a dihedral
group $D_4$, with eight elements.  Thus, the topology of the spacetime
is $\R^4/D_4$, and there is an isolated $D_4$ singularity at the origin.

How could
 this metric be obtained from a detailed
calculation, rather than being deduced from the topology as we have
done?  In supergravity, the metric arises by evaluating
 the relevant terms in the
supergravity action such as $\int H^2$.  The correction to the Bianchi identity
\niid\  will modify $H$ and therefore modify the evaluation of the
metric coming from this term; this correction is also related
by supersymmetry to additional terms in the action, which will likewise
enter
in computing explicitly the metric.  Taking all these effects
into account, one could in principle
generate a string perturbation expansion which must add up to \zzzz,
since it is determined by the symmetries, the hyper-Kahler structure,
and the one-loop effect that determines the topology.

\subsec{Exact Metric}

In determining the metric to all finite orders in $\alpha'$,
we have used a symmetry under shifts of the $B$-field period.
This symmetry is violated by worldsheet instantons wrapped on the
exceptional divisor $S$.  Moreover \witteno, such instantons do correct
the metric on $\M$, because there are no worldsheet fermion zero modes
except the minimal set required by supersymmetry.   (In fact, the normal
bundle to $S$ is ${\cal O}(-2)$, and the gauge bundle is trivial; eqn.
(3.2) of \witteno\ is thus applicable and shows that the instanton 
contribution  is not zero.) 
The instanton contributions vanish exponentially
fast at infinity
on $\M$ (since they are proportional to $\exp(-{A/2\pi \alpha'})$
with $A$ the area of $S$).  So we want a hyper-Kahler metric 
that has $SU(2)_R$ symmetry, rotating the complex structures,
and is exponentially close to \zzzz\ at infinity.

Four-dimensional hyper-Kahler metrics with such
an $SU(2)_R$ symmetry and no $U(1)_A$ symmetry
have been classified in \ah. 
A smooth metric of this type exists if and only if the first
Chern class at infinity is $-4$ or $-2$ (or $+4$ or $+2$ if one takes
the opposite orientation on the fiber, which corresponds to starting
with the opposite hyper-Kahler structure on $(\R^3\times \S^1)/\Z_2$). 
The smooth manifold with first Chern class $-4$ is often
called the Atiyah-Hitchin space ${\cal M}_{AH}$; it is the moduli
space of BPS dimonopoles on $\R^3$.  The fundamental group of ${\cal M}_{AH}$
is $\Z_2$; it therefore has a double cover, which is the smooth manifold
with first Chern class at infinity $-2$.

The fact that the first Chern class at infinity that we computed
(namely $-4$) corresponds to one of the values leading to a smooth
manifold suggests that the moduli space we want is in fact ${\cal M}_{AH}$.
Can we argue {\it a priori} that ${\cal M}$ should be smooth?

In the introduction, we argued that near an orbifold singularity without
small instantons, the effective heterotic string coupling 
(if small at infinity) is uniformly
small, so that nonperturbative effects should not arise.
For example, this means that ${\cal M}$ should not have singularities
interpreted in terms of nonperturbative massless particles or
a non-trivial infrared CFT.
Any singularity in ${\cal M}$ must make sense from the point of view
of conformal field theory.

Conformal field theory corresponds to the tree approximation to string
theory, so this means that any singularity
in ${\cal M}$ should have an interpretation in the tree approximation
to a weakly coupled classical field theory.  Moreover (since we are
looking at the heterotic string on a hyper-Kahler four-manifold),
this must be a supersymmetric field theory in six dimensions.
This is very restrictive: in this framework, the only mechanism to
generate a singularity is via un-Higgsing of a gauge symmetry.
For example, an orbifold singularity could be interpreted in classical
field theory in terms of restoration of a discrete gauge symmetry.
A $\Z_2$ orbifold singularity can likewise be interpreted in terms
of un-Higgsing of a $U(1)$ gauge symmetry as in the hyper-Kahler
quotient considered in \tolgo, or in terms of un-Higgsing of an
$SU(2)$ symmetry using a construction discussed in 
\ref\seiwitfam{N. Seiberg and E. Witten, ``Monopoles, Duality, And
Chiral Symmetry Breaking In ${\cal N}=2$ Supersymmetric QCD,'' 
hep-th/9408099.}.

Thus, the orbifold singularities of the supergravity moduli space
${\cal M}_{SUGRA}=(\R^3\times \S^1)/\Z_2$ could be interpreted
in classical supersymmetric field theory.  String perturbation theory
corrects this to a moduli space ${\cal M}_{\alpha'}$ with a metric
given in \zzzz.  
Now, instead of two $A_1$ singularities, there is a single $D_4$
singularity.  Again, this could be interpreted in terms of classical
field theory in terms of restoration of either a discrete gauge
symmetry (the dihedral symmetry $D_4$) or a continuous gauge symmetry
(the gauge symmetry used \refs{\kron,\dougmoore} in interpreting the
$D_4$ singularity as a hyper-Kahler quotient).

However, since the worldsheet instanton corrections to the metric
are nonzero, this is not an option for the description of the moduli
space.  The conformal field theory moduli space ${\cal M}$ has
the $SU(2)_R$ symmetry with three-dimensional orbits, but no
$U(1)_A$ symmetry.  We can now use the analysis in chapter 9 of \ah.
Hyper-Kahler metrics with the asymptotic behavior we want and no $U(1)_A$
symmetry correspond
to trajectories that flow to the point labeled $Q$ in Diagram 4 of that
chapter
and do not lie on the line $QB$.   It is shown in \ah\ that hyper-Kahler
manifolds 
obtained this way are either $\M_{AH}$ or its double cover, or have
a singularity in real codimension one where the trajectory originates
at $B$.  Such a real codimension one singularity cannot arise from
a hyper-Kahler quotient, and so could not be interpreted in
weakly coupled supersymmetric field theory.  Given this, the arguments
in the introduction plus the nonvanishing of the instanton corrections
imply ${\cal M}={\cal M}_{AH}$.

\subsec{Linear Sigma Model Approach}

Smoothness of the moduli space ${\cal M}$ presumably means that
the conformal field theory describing the heterotic string at
an $A_1$ singularity, without small instantons, is uniformly
valid for small string coupling constant.  This contrasts with
the case of a small instanton, where the effective string coupling
diverges and nonperturbative phenomena occur no matter how weak
the bare string coupling might be.

The arguments in the introduction suggest that more generally,
a singularity in the metric with no singularity of the gauge field
tends not to cause a breakdown of heterotic string perturbation theory.
We will here give a simple linear sigma model argument that supports
this expectation for a large class of examples.  In making the analysis,
we will adopt the proposal in \nref\ugwitten{E. Witten, ``Phases
Of $N=2$ Theories In Two Dimensions,'' Nucl. Phys. {\bf B403} (1993) 
159, hep-th/9301042.}%
\nref\tugwitten{E. Witten, ``Some Comments On String Dynamics,'' in
{\it Strings '95: Future Perspectives In String Theory}, ed. I. Bars et. al.,
hep-th/9507121.}%
\refs{\ugwitten,\tugwitten}
according to which a breakdown of conformal field theory should
be detected by a failure of normalizability when the quantum states
``spread'' in a new direction in field space.

We consider  the heterotic string on
an $n$-dimensional complex hypersurface $K$ defined by an equation
\eqn\uxni{F(\phi_1,\phi_2,\dots,\phi_{n+1})=0}
in $n+1$ complex variables $\phi_1,\phi_2,\dots,\phi_{n+1}$.
$K$ is smooth if the equations
\eqn\buxin{0=F={\partial F\over \partial \phi^i}}
have no common solution.  $K$ can be regarded as a noncompact
Calabi-Yau manifold.  We want to consider what happens when,
by varying a parameter, a singularity develops.  For example, we may take
\eqn\cuxin{F=\sum_{i=1}^{n+1}\phi_i^2+\epsilon,}
with a parameter $\epsilon$.  In this case, the hypersurface $K$ develops
a ``conifold'' singularity as $\epsilon\to 0$.  
In studying the perturbative heterotic string near
a singularity in $K$, we will assume that the gauge bundle is
trivial.  This means that the left-moving
gauge fermions of the heterotic string will be free fields, decoupled
from the $(0,2)$ or $(0,4)$ 
 sigma model that will describe the motion of strings on
$K$.

Indeed, if $n>2$, the sigma model with target $K$ is a $(0,2)$ model,
while for $n=2$, it is a $(0,4)$ model.  Related to this, if
 $n>2$, the only moduli of such a singularity are the complex
structure moduli that are present in $K$. For $n=2$, because there are collapsing two-spheres at the singularity, there are additional Kahler and $B$-field moduli. 

\nref\dista{J. Distler, B. R. Greene, and D. R. Morrison,
``Resolving Singularities in (0,2) Moduli Space,'' hep-th/9605222.}%
\nref\distb{T.-M. Chiang, J. Distler, and B. R. Greene, ``Some Features
Of (0,2) Moduli Space,'' hep-th/9702030.}%
\ref\silver{E. Silverstein
and E. Witten, ``Criteria For Conformal Invariance of $(0,2)$ Models,''
hep-th/9503212.}%
For $n>2$, I will argue  using $(0,2)$ linear sigma
models that the conformal field theory remains nonsingular as $K$ develops
a singularity.\foot{Arguments along these lines have
been developed in detail in
\refs{\dista,\distb}, and some aspects were explored
in \silver.}
   For $n=2$ (which is the case most directly relevant to the
present paper) it is difficult to construct an equally satisfactory
$(0,4)$ linear sigma model, and the argument based on $(0,2)$
linear sigma models is less satisfactory because it does not
exhibit all of the moduli.  But I believe the result is still
true for $n=2$.\foot{$(0,4)$ linear sigma models were
constructed in \dougmoore\ by considering $D1$-brane probes of
ADE singularities.  Such models often have gauge anomalies, which
were interpreted in \dougmoore\ via anomaly inflow to the probe
from the bulk of spacetime; but this interpretation
does not seem relevant for our consideration of heterotic string
conformal field theory.  An anomaly-free $(0,4)$ linear sigma model
with exactly the properties we would want to exhibit all the moduli
and establish our claim for $n=2$ does not seem to exist.  An oversight
in the original version of this paper (where an anomalous model
was considered) was pointed out by M. Aganagic
and A. Karch; I thank them and A. Mikhailov for discussions.}

To construct a linear sigma model that should flow in the infrared
to the $(0,2)$ superconformal
field theory with target space $K$, we work in $(0,2)$ superspace
(described more fully in \ref\wittenhull{C. Hull and E. Witten,
``Supersymmetric Sigma Models And The Heterotic String,'' Phys. Lett.
{\bf 160B} (1985) 398.})
with supercovariant derivatives $D_+,\bar D_+$ obeying $D_+^2={\bar D}_+^2
=0$,  $\{D_+,\bar D_+\}=\partial_+$.  
To get $(0,2)$ supersymmetry, $D_+$ and $\bar D_+$ both have a spinor
index of the same chirality, as do the fermionic coordinates $\theta^+$,
$\bar \theta^+$ of superspace.
We introduce bosonic chiral superfields
$\Phi_1,\dots,\Phi_{n+1}$, obeying $\bar D_+\Phi_i=0$.  They can
be expanded
$\Phi_i=\phi_i+i\theta^+\psi_i^-+\dots$, with $\phi_i$ and $\psi_i^-$
complex bosonic and fermionic fields, respectively; $\psi_i^-$ is of
definite chirality.  The conventional
kinetic energy for these fields is contained in the superspace
expression
\eqn\offgo{L_{kin}=\int d^2\sigma d^2\theta^+ \sum_i\bar\Phi_i
\partial_-\Phi_i.}
In addition, we introduce a fermionic chiral superfield $\Lambda^+= 
\lambda^++\theta^+p+\dots$, where $\lambda^+$ is a complex fermion of 
opposite
chirality to $\psi_i^-$, and $p$ is a complex bosonic auxiliary field.
The kinetic energy for $\lambda^+$, together with a $|p|^2$ term,
comes from
\eqn\laux{L_{aux}=\int d^2\sigma d^2\theta^+ \bar\Lambda^+\Lambda^+.}
These multiplets are coupled by a ``superpotential'' interaction
\eqn\baux{L_{super}=\int d^2\sigma d\theta^+\Lambda^+ F(\Phi_i)+{c.c.}}
Note that the integrand must be a chiral superfield in order for 
\baux\ to be supersymmetric; that is why $F$ must be holomorphic.  We will
assume moreover that $F$ is a polynomial, so that the superrenormalizable
quantum field
theory we are constructing exists rigorously, and the
only issue is what it flows to in the infrared.
After performing the $\theta$ integral, $L_{super}$ gives an
interaction $pF(\phi_i)$ (plus a Yukawa coupling that gives mass to
$\lambda^+$ together with a $\phi$-dependent
linear combination of the $\psi_i^-$).
After integrating out the auxiliary field $p$ using the $|p|^2$ term
from $L_{aux}$, we get then an
ordinary potential
\eqn\tobbo{V=|F(\phi_i)|^2.}

The space of classical zeroes of $V$ is thus the hypersurface $K$ defined
by $F=0$.  If $K$ is smooth, this model       presumably flows
to a $(0,2)$ superconformal field theory with that target.
Even if $K$ is not smooth, as long as its singularities are isolated,
the possible occurrence of a singularity in $K$ should not affect the
well-definedness of this conformal field theory.  For example, in this
theory, since $F$ (being a polynomial) grows if one is far from $K$,
the wave functions decay rapidly when far away from $K$, whether $K$
is singular or not.  The only unnormalizability of the quantum states
comes from the noncompactness of $K$, and assuming the singular
set of $K$ is compact (by holomorphy this is so precisely if the
singularities of $K$ are isolated) this unnormalizability is not affected
by the singularities.  When $K$ develops a singularity, the conformal
field theory becomes strongly coupled and difficult to analyze
near the singularity, but nonetheless should continue to be well-behaved.

\bigskip\noindent{\it $(2,2)$ Models}

Since the above arguments may appear to be based on almost nothing, let
us now show that in fact a similar analysis with a singularity in
the gauge bundle as well as in the geometry gives a different
result.  We will consider the special case of ``embedding the
spin connection in the gauge group,'' which ensures that the gauge bundle becomes singular when the geometry does.
   To study this case, we must
consider $(2,2)$ superconformal field theories, and formulate
our linear sigma models in $(2,2)$ superspace.  This was done for the 
conifold in \ugwitten, but here we will follow a more elementary and
direct route.  We introduce bosonic
chiral superfields $\Phi_i=\phi_i+\dots
$ and ${\cal P}=P+\dots$
and a superpotential
\eqn\inci{W={\cal P}F(\Phi_i).}
The ordinary potential is as usual in $(2,2)$ models $|dW|^2$,
which in this case gives
\eqn\binci{V=|F|^2+|P|^2\sum_i\left|{\partial F\over 
\partial \phi_i}\right|^2.}
A classical zero of $V$                with $P\not=0$ must
have $F=dF=0$.  Hence, if the hypersurface $K$ is smooth, all zero
energy states have $P=0$.
The space of such states is the hypersurface $K$ obtained by setting
$F=0$ in ${\bf C}^{n+1}$.
Thus, for smooth $K$, the model should flow in the infrared to a sigma
model with target space $K$.  So far this discussion is rather like
the $(0,2)$ case.  But now suppose that $K$ is singular.  Setting
the $\phi_i$ to a singular point of $K$, that is a solution of \buxin,
we now get a new branch of the moduli space of vacua by taking $P\not= 0$.
The new branch is not compact, since $P$ can be arbitrarily big.
The ability of quantum states to spread on this new branch should
be expected to lead to a breakdown of the conformal field theory.
Indeed, in the case of the conifold (for $n=3$), the familiar
pole in Yukawa couplings has been computed \silver\ from the ``leaking'' of quantum states onto
the new branch.  This computation was actually done in a 
linear sigma model realization of the conifold different
from what we have given above; the fact that   different
linear sigma model formulations show the occurrence of a new branch
at the singularity encourages
us to believe that this
is an intrinsic phenomenon of the singularity
and not an artifact of a particular linear sigma model formulation.

\bigskip\noindent{\it Comparison With Bundle Singularities}

The opposite of the situation we have just looked at is a singularity
in the gauge bundle on a smooth manifold.  
It has been argued (\silver, section 5.1) 
that in a large class of linear sigma models,
a gauge singularity on a smooth manifold does result in
a breakdown of conformal field theory.  For a particular
case (in complex dimension three) a proposal has been made
concerning the nature of the resulting nonperturbative physics
\ref\seisilka{S. Kachru, N. Seiberg, and E. 
Silverstein, ``SUSY Gauge Dynamics And Singularities of $4$-$D$ ${\cal N}=1$
String Vacua,'' Nucl. Phys. {\bf B480} (1996) 170, hep-th/9605036.}. 
These results generalize
the small instanton story for $n=2$.  When conformal
field theory does break down because of a singularity, the nature
of the breakdown is not fully understood.
In the   known cases, including the small instanton
\chs, the Type II 
$A$-$D$-$E$ singularities \ref\vafaooguri{C. Vafa and H. Ooguri,
``Two-Dimensional Black Hole And Singularities Of CY Manifolds,''
Nucl. Phys. {\bf B463} (1996) 55, hep-th/9511164.},
and several cases treated recently 
\nref\latesw{N. Seiberg and E. Witten, ``The $D1/D5$ System And
Singular CFT,'' JHEP {\bf 9904:017} (1999), hep-th/9903224.}%
\nref\kg{A. Giveon and D. Kutasov, and O. Pelc, ``Holography For Noncritical
Superstrings,'' hep-th/9907178.}%
\refs{\latesw,\kg}, the breakdown of conformal field theory can apparently
be described by the appearance, after suitable change of variables,
of a linear dilaton field with a blowup of the string coupling constant
at one end.  A framework for understanding this has been proposed
\ref\ahab{O. Aharony and M. Berkooz, 
``IR Dynamics of $d=2$, ${\cal N}=(4,4)$ Gauge Theories
and DLCQ of `Little String Theories,' '' hep-th/9909101.}.

\appendix{}{ Structure Of $(0,2)$ Moduli Space}

Let $Y$ be a Calabi-Yau threefold; keep its complex
structure fixed in this discussion.  The remaining moduli of a $(0,2)$
supersymmetric model with target $Y$ are the gauge bundle moduli and the
Kahler moduli.  The definition of a holomorphic vector bundle depends
only on the complex structure and not the Kahler metric.
One might therefore think that in the supergravity approximation,
the $(0,2)$ 
moduli space would be a product ${\cal M}_G\times {\cal M}_K$ of a gauge 
bundle moduli space ${\cal M}_G$
and a Kahler moduli space ${\cal M}_K$.  This was assumed in section
5.2 of \silver, but is inaccurate for several reasons. (The inaccuracy
has been corrected in the hep-th version of the paper.)
One reason is that ${\cal M}_G$ is the moduli space of {\it stable}
bundles, and the condition for stability ``jumps'' as the Kahler
metric varies.   As a result, ${\cal M}_G$ undergoes birational
transformations as one moves about in ${\cal M}_K$, a phenomenon
explored in \ref\sharpe{E. Sharpe, ``Kahler Cone Substructure,''
hep-th/9810064.}.
There is a reciprocal effect which, by itself, would cause
 ${\cal M}_K$ to be fibered over ${\cal M}_G$.
Indeed, given that the $B$-field periods are part of the Kahler multiplets,
a non-trivial fibration of ${\cal M}_K$ over ${\cal M}_G$ follows
from  the fact developed in \witteno\ and used in section 2.2 above:
the $B$-field periods take values in a circle bundle, not just in $U(1)$.
(The nontrivial fibration of the $B$-field periods over ${\cal M}_G$ comes
from the $\tr \,F\wedge F$ term in the Bianchi identity for $B$; this
term is present in the minimal supergravity.)

Regrettably, this invalidates the attempt made originally
in section 5.2 of
\silver\ to argue conformal invariance of (0,2) models directly
from nonlinear sigma models.    (The rest of the paper
is based on quite different arguments using linear sigma models.)
Note that the formula for the instanton contribution to the superpotential
given in \witteno\ varies holomorphically with the parameters,
showing that, because of the nontrivial fibration, it is possible
for the contribution of a given instanton to make a nonvanishing
contribution to the superpotential that obeys all conditions
of holomorphy.
\bigskip

I would like to thank H. Ooguri for helpful comments.
\bigskip
This work was supported in part by NSF Grant PHY-9513835 and by the
Caltech Discovery Fund.  
\listrefs
\end